# Broadband Dispersive-Wave Emission Coupled with Two-Stage Soliton Self-Compression in Gas-Filled Anti-Resonant Hollow-Core Fibers

J. Y. Pan,[1,3,4,§] Z. Y. Huang,[1,§,*] Y. F. Chen,[1] F. Yu,[4,5] D. K. Wu,[4,5] T. D. Chen,[1,3] D. H. Liu,[1,3] Y. Yu,[1,3,4] Xin Jiang,[2] M. Pang,[1,2,4,†] Y. X. Leng,[1,4,‡] and R. X. Li[1]

[1]State Key Laboratory of High Field Laser Physics and CAS Center for Excellence in Ultra-intense Laser Science, Shanghai Institute of Optics and Fine Mechanics (SIOM), Chinese Academy of Sciences (CAS), Shanghai 201800, China

[2]Russell Centre for Advanced Lightwave Science, Shanghai Institute of Optics and Fine Mechanics and Hangzhou Institute of Optics and Fine Mechanics, Hangzhou, 311421, China

[3]Center of Materials Science and Optoelectronics Engineering, University of Chinese Academy of Sciences, Beijing 100049, China

[4]Hangzhou Institute for Advanced Study, Chinese Academy of Sciences, Hangzhou 310024, China

[5]Key Laboratory of Materials for High Power Laser, Shanghai Institute of Optics and Fine Mechanics, Chinese Academy of Sciences, Shanghai 201800, China

We studied the underlying mechanism of broadband dispersive-wave emission within a resonance band of gas-filled anti-resonant hollow-core fiber. Both theoretical and experimental results unveiled that the high-order soliton, launched into the hollow-core fiber, experienced two stages of pulse compression, resulting in a multi-peak structure of the dispersive-wave spectrum. Over the first-stage pulse compression, a sharp increase of the pulse peak power triggered the first time of dispersion-wave emission, and simultaneously caused ionization of the noble gas filled in the fiber core. Strong soliton-plasma interactions led to blue shifting of the pump pulse, and the blue-shifted pulse experienced a decreasing dispersion value in the fiber waveguide, resulting in an increase of its soliton order. Then, the second-stage pulse compression due to the high-order soliton effect triggered the second time of dispersive-wave emission at a phase-matched frequency slightly lower than that in the first stage. Multi-peak spectra of the output dispersive-waves and their formation dynamics were clearly observed in our experiments, which can be understood using a delicate coupling mechanism among three nonlinear effects including high-order-soliton compression, soliton-plasma interaction and phase-matched dispersive-wave emission. The output broadband dispersive-wave could be potentially compressed to sub-30 fs duration using precise chirp-compensation technique.

Anti-resonant hollow-core fiber (AR-HCF) [1-4], with broad transmission band and tightly-confined optical fields in its hollow-channel core, has been widely regarded as an ideal platform for studying light-gas interactions [5-7]. Two remarkable features of a gas-filled hollow-core fiber system [5,6], including the elongated interaction length over fiber waveguide and tunable dispersion landscape through gas type and pressure controls, render many intriguing applications from ultrafast pulse compression [8,9] to high-efficiency laser frequency conversion [10-14]. Of particular importance, the AR-HCF provides distinctive dispersion properties over its res[1]onance bands [15-17]. While the resonance-type energy coupling between the core and cladding modes leads to high fiber losses, the refractive index of the fundamental optical mode of the fiber varies intensely within the resonance band, resulting in strongly-modulated dispersion values. This unique mechanism of fiber dispersion manipulation [15-17], being dependent solely on the capillary thickness of the resonance element [15-17], enables several important applications, such as the generation of multi-octave supercontinuum light [18] and high-efficiency laser energy conversion to some specified frequencies [19-23].

In previous studies, the theory of phase-matching between the nonlinear pump pulse and a packet of quasi-linear waves [18-22] was successfully used to explain the narrow-band, high-efficiency dispersive-wave (DW) generation within the resonance band of the AR-HCF. Spectral broadening phenomena of these in-resonance DWs [24,25] at higher pulse energies, as well as the appearance of multi-peak structures on their spectra

*huangzhiyuan@siom.ac.cn

†pangmeng@siom.ac.cn

‡lengyuxin@siom.ac.cn

§These authors contributed equally to this work.

[24,25] are still obscure in physics so far. Here, we unveil that the formation of these broadband, structured DW spectra can be attributed to the coupling mechanism among three nonlinear optical effects (self-compression of high-order soliton, soliton-plasma interaction and phase-matched DW emission). Two pulse-compression stages were found in our studies, responsible to the generation of multi-peak structures on the DW spectra. We demonstrate that the phase matching between the quasi-linear DW and the blue-shifting soliton [5-7,10,11] can predict precisely the spectral positions of the multi-peak DW components. Moreover, frequency-resolved optical gating (FROG) measurements were performed in experiments, pointing out the possibility of sub-30-fs DW pulse generation after precise chirp compensation.

As illustrated in Fig. 1(a), in the experiment ultrafast pump light was launched into an 18-cm length of AR-HCF, filled with 1.2-bar argon gas. The pump light has a pulse repetition rate of 1 kHz, a central wavelength (central frequency) of 775 nm (386 THz), and a full-width-half-maximum pulse width of ~19 fs, which was generated from a commercial ultrafast Ti: Sapphire laser (1 kHz, 45 fs, 4.5 mJ, 800 nm) and a home-made pulse compressor based on a section of gas-filled capillary fiber and several chirp mirrors [11,23-25]. Before launching into the AR-HCF, the pump pulse was measured using a second-harmonic-generation (SHG) FROG set-up, see Figs. 1(b) and 1(c) for the FROG results. The AR-HCF used in the system has a core diameter of ~24 μm and a capillary-wall thickness of ~260 nm. The scanning-electron microscopy image of the fiber is illustrated in the insert of Fig. 1(d). When 1.2-bar Ar gas was filled into the AR-HCF, the group-velocity dispersion curve of the fiber fundamental ($HE_{11}$) optical mode can be calculated using the analytical model demonstrated by Zeisberger and Schmidt [15], and the results are plotted as the solid blue lines in Fig. 1(d). Using the bouncing-ray model [17], the loss curve was also calculated and the results are plotted as the solid purple line in Fig. 1(d). The cyan shadow in Fig. 1(d) indicates one resonance band of the AR-HCF at around 539 THz (556 nm).

In the experiment we gradually increased the pump pulse energy from 1 μJ to 3.9 μJ (an increase of the soliton order from 1.2 to 2.4), and measured the output spectra as plotted in Fig. 1(e). It can be found that as the pulse energy increased, the output spectrum was first broadened due to the high-order soliton effect [26]. The phenomenon of soliton blue-shifting was observed at pump pulse energies higher than ~2 μJ, accompanied by the spectral broadening of the in-resonance DW emission, as illustrated in Fig. 1(e). A typical output light spectrum from the AR-HCF, at a pump pulse energy of 3.9 μJ, is exhibited in Fig. 1(f). The broadband DW at ~530 THz (~565 nm) has a high spectral ratio of ~38%, corresponding to a total frequency-conversion efficiency of ~13% (calculated through dividing the energy of the output DW by the pump pulse energy).

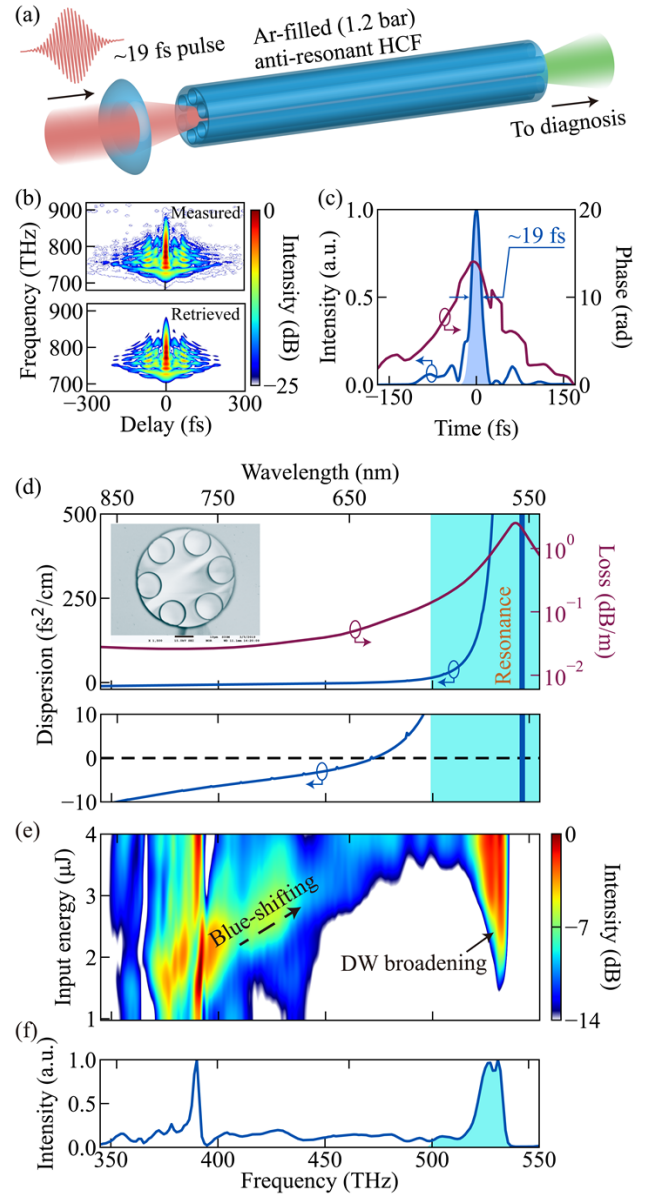

FIG. 1. (a) Conceptual scheme of the experiment. (b) Measured and retrieved SHG-FROG traces of the pump pulse. (c) Retrieved temporal profile (solid blue line) and phase (solid purple line) with considering the dispersion introduced by the 1.5-mm-thick fused silica window. The blue shadow represents the effective energy after removing the pulse pedestals,

accounting for ~64% of the total pulse energy. (d) Simulated dispersion (solid blue line) and loss (solid purple line) of the AR-HCF filled with 1.2 bar Ar. The cyan shadow indicates the resonant spectral region. The inset represents the scanning electron micrograph of the AR-HCF. (e) Measured spectral evolutions out of the AR-HCF as a function of input pulse energy. The dashed black arrow points out the soliton blue-shifting due to gas ionization. (f) The measured output spectrum at a pump energy of 3.9 μJ. The cyan shadow represents the broadband DW spectrum.

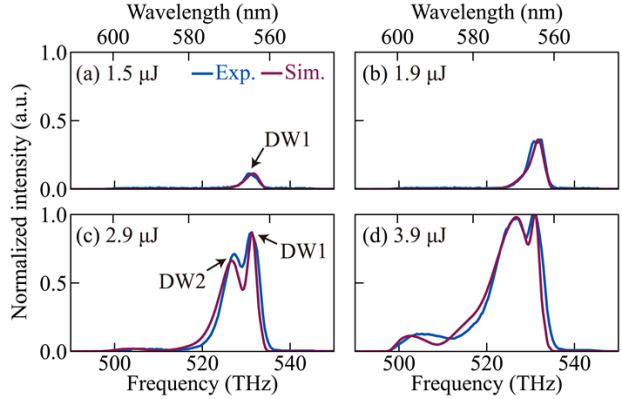

FIG. 2. Experimental (solid blue line) and simulated (solid purple line) spectra of DW output from the AR-HCF at several discrete input pulse energies of (a) 1.5 μJ, (b) 1.9 μJ, (c) 2.9 μJ and (d) 3.9 μJ. The spectra in (a)-(c) are normalized based on the spectral intensity in (d). The black letters "DW1" and "DW2" in (a) and (c) indicate the spectral peaks of dispersive waves.

To illustrate the generation dynamics of the broadband DW, a short-wavelength-pass optical filter with a cut-off wavelength of 600 nm was used in the experiment to filter out the DW from the output light spectrum. Four typical spectra at different pump pulse energies are exhibited in Fig. 2. While these measured DW spectra are plotted as solid blue lines, numerical results simulated using the single-mode unidirectional pulse propagation equation (UPPE) model [27,28] are as the solid purple lines in Fig. 2. In the numerical simulation, we used the 19-fs pulse as the pump pulse whose profile was measured by the SHG-FROG set-up, see Figs. 1(b) and 1(c). The coupling efficiency of the pump light to the AR-HCF was set to be ~0.9. It can be found in Fig. 2(a) that at a relatively-low pulse energy (1.5 μJ), a narrow-band DW component started to appear at ~565 nm (~530 THz), which can be well understood using the theory of phase matching [24,29]. The intensity of this narrow-band DW increased as the pump pulse energy increased [see Fig. 2(b)], and at a pump energy of 2.9 μJ the second DW component at ~568 nm (~527 THz) was observed on the spectrum [see Fig. 2(c)]. As the pump pulse energy increased further to 3.9 μJ, this second spectral component continued to grow and simultaneously its spectrum became wider, leading to the generation of broadband, multi-peak structures on the DW spectrum, see Fig. 2(d). As shown in Fig. 2, the UPPE model can be used to reproduce the DW generation process and the simulation results exhibit striking agreements with the experimental results.

In order to fully understand the underlying mechanism concerning nonlinear pulse propagation and multi-peak DW generation, we illustrated in Fig. 3 the evolution of optical fields inside the 25-cm-long AR-HCF at a pump pulse energy of 3.9 μJ. The simulation results were obtained using the UPPE model [27,28] in which the plasma density inside the AR-HCF core, due to ionization of the Ar gas, was calculated using the Perelomov-Popov-Terent'ev model [30].

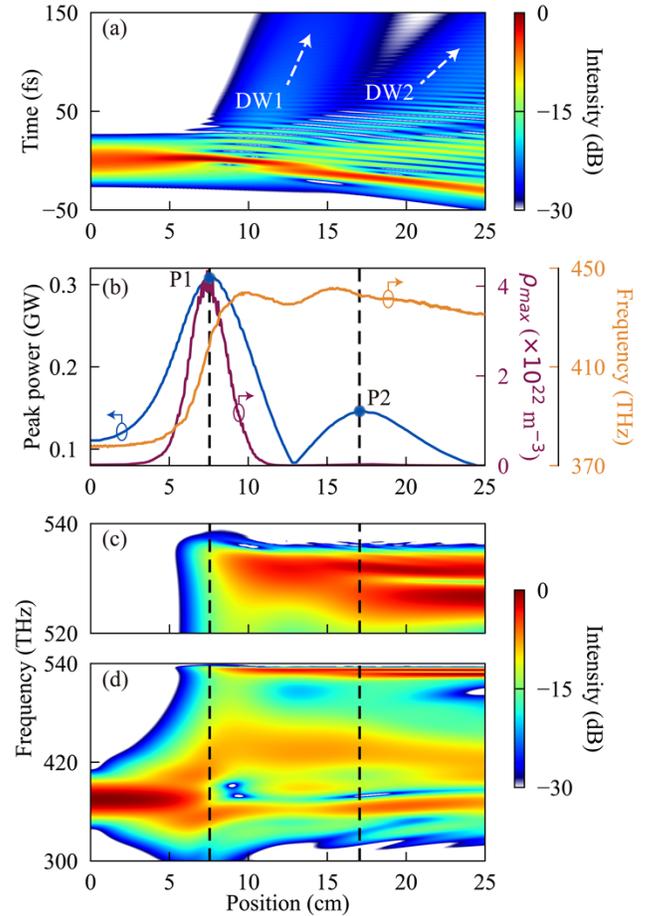

FIG. 3. Simulated temporal (a) and spectral (d) evolutions of the pulse in a 25-cm-long Ar-filled AR-HCF at the gas pressure of 1.2 bar and pump pulse energy of 3.9 μJ. The dashed white arrows indicate the radiation direction of the two DWs. (b) The peak power (solid blue line), maximum plasma density (solid

purple line) and central frequency (solid orange line) of the soliton pulses at different positions in fiber. (c) The spectral evolution of DW magnified from panel (d). The two dashed black lines in (b)-(d) represent the fiber positions P1 and P2 when the two self-compression of the soliton pulses reach the maximum temporal compression point.

At a pulse energy of 3.9 μJ (a soliton order of 2.4), the pump pulse, after launched into the AR-HCF, first experienced self-compression due to the high-order soliton effect [26], leading to a sharp increase of its peak intensity, see Fig. 3(a). The pulse peak power reached the first maximum at the fiber position of 7.55 cm [see Fig. 3(b)], accompanied by a dramatic broadening of the pulse spectrum [see Figs. 3(c) and 3(d)]. This led to the first time of phase-matched DW emission at a frequency of ~533 THz (~562 nm), marked as DW1 in Fig. 3(a).

Over the first-stage pulse compression (from 0 to 7.55 cm in the AR-HCF), the light field intensity exceeded ionization threshold of the Ar gas, and the gas-ionization process consumed some of the pulse energy [5,7,10]. After the maximum compression point, both the pulse-width increasing due to the high-order-soliton breathing [26] and the pulse-energy decreasing due to the gas ionization resulted in a quick drop of the pulse peak power and therefore a quick decrease of the plasma density, see Figs. 3(a) and 3(b). During the accumulation of the plasma density, strong light-plasma interactions shifted the central frequency of the soliton to the blue (high-frequency) side (from ~375 THz to ~440 THz) [5,7,10], which is also illustrated in Fig. 3(b). At higher frequencies, the optical pulse experienced a decreasing absolute value of group-velocity dispersion in the AR-HCF [see Fig. 1(d)], counteracting the ionization-induced pulse-energy decreasing. Therefore, the soliton order was maintained to be a relatively-high value during this complex nonlinear process. The second-stage of pulse compression was then observed due to the temporal-breathing effect of the high-order soliton [26], resulting in the re-rising of the pulse peak power over the fiber length from ~13 to ~17 cm, see Fig. 3(b).

The re-compressed pulse had a high peak power at the fiber position of 17.05 cm (P2) [see Fig. 3(b)], leading to the second DW emission with a phase-matched frequency of ~527 THz (568 nm), marked as DW2 in Figs. 3(a). The two different phased-matched frequencies of DW emission processes can be well understood using the theory of phase matching, expressed as [24,29]:

$$\Delta\beta(\omega) = \beta(\omega) - \beta_0 - \beta_1(\omega - \omega_0) - \beta_{Kerr} - \beta_{Plasma}, \quad (1)$$

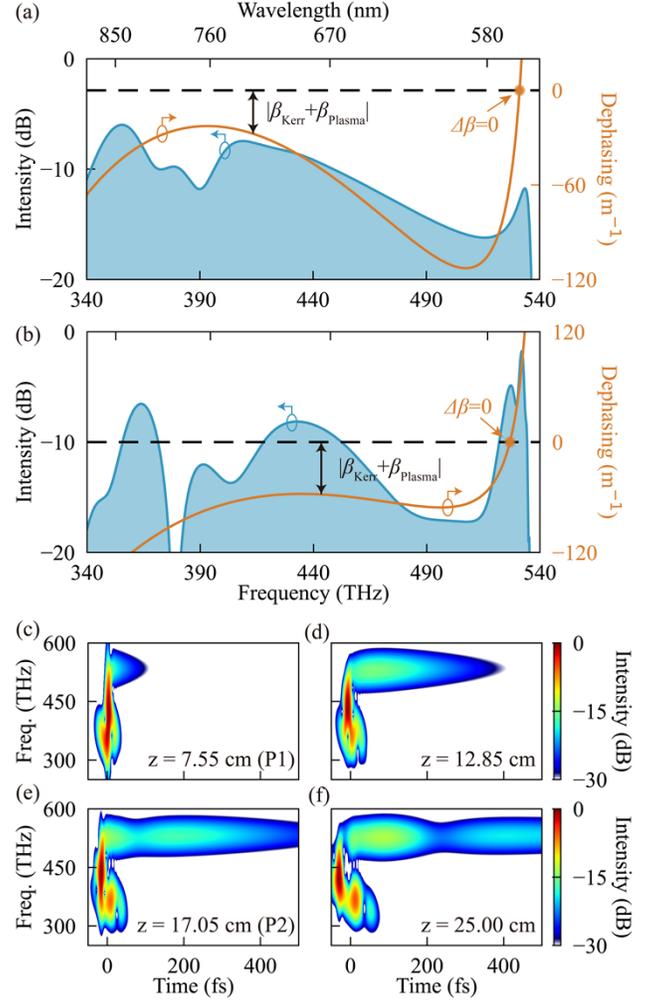

FIG. 4. (a) and (b) Simulated spectral profiles (shaded blue areas) of the pulses at different positions P1 and P2 shown in Fig. 3 and the corresponding dephasing (solid orange lines) of DWs. Phase-matching DW emission is achieved at zero dephasing (solid orange circles). The bidirectional black arrows represent the phase mismatch at soliton central frequency defined as the spectral centroid of the soliton pulses without considering the Kerr effect and plasma contribution. Time-frequency analysis of the pulses at different positions of (c) 7.55 cm (P1), (d) 12.85 cm, (e) 17.05 cm (P2) and (f) 25.00 cm, calculated using a 5-fs Gaussian gate pulse.

where $\Delta\beta(\omega)$ is the propagation-constant difference between the solitonic wave and the quasi-linear DW, $\beta(\omega)$ the propagation constant of the DW. While $\beta_0$ and $\beta_1$ are zero- and first-order derivations of the propagation constant related to the central frequency of the soliton ($\omega_0$), $\beta_{Kerr}$ and $\beta_{Plasma}$ are the contributions of Kerr and plasma effects on the propagation constant of the soliton.

These two terms can be expressed as $\beta_{\text{Kerr}} = \gamma P_{\text{P}} \frac{\omega}{\omega_0}$ and $\beta_{\text{Plasma}} = -\frac{\omega_0}{2n_0 c}\frac{\rho}{\rho_{\text{cr}}}\frac{\omega_0}{\omega}$ [24,29], where $\gamma$ is the Kerr-nonlinearity coefficient, $P_{\text{P}}$ the peak power of the self-compressed soliton, $n_0$ the refractive index at $\omega_0$, $c$ the speed of light in vacuum, $\rho$ the plasma density at the DW-generation position, and $\rho_{\text{cr}}$ the critical free-electron density of Ar gas.

The simulated light spectra at the two DW emission positions [P1 and P2 in Fig. 3(b)] are illustrated as blue shadows in Figs. 4(a) and 4(b), while the propagation-constant-difference curves ($\Delta\beta(\omega)$) as the function of optical frequency ($\omega$) for the two cases are plotted as the orange lines in these two figures. It can be found that at the position of 7.55 cm [see Fig. 4(a)], the soliton central frequency ($\omega_0$) was slightly shifted to ~413 THz due to soliton-plasma interactions, giving rise to a phased-matched ($\Delta\beta = 0$) frequency of ~531 THz. At the second self-compression position [17.05 cm, see Fig. 4(b)], the soliton central frequency was strongly shifted to ~443 THz, giving rise to a lower phase-matched frequency of the DW (~527 THz). The off-set of $\Delta\beta(\omega)$ curve at the soliton central frequency for both the two cases is due to the nonlinear terms of $\beta_{\text{Kerr}}$ and $\beta_{\text{Plasma}}$.

Time-frequency analysis of the optical field at four different fiber positions (7.55 cm, 12.85 cm, 17.05 cm and 25.00 cm) were also performed in our simulations and the results are illustrated in Figs. 4(c)-4(f). Two stages of DW emission with slightly-different DW frequencies were clearly observed in these holograms, and this two-stage-emission mechanism resulted in an obvious time delay between the two packets of DWs, see Fig. 4(f).

At a pump pulse energy of 3.9 μJ, we performed SHG-FROG measurement on the output DW filtered by the short-wavelength-pass filter. The measured FROG signal [see Fig. 5(a)] shows good agreement with the retrieved signal [see Fig. 5(b)]. Using the SHG-FROG signal, we obtained the full spectral-temporal information of the output DW, the results are illustrated in Figs. 5(c)-5(e). While the retrieved temporal profile of the DW and its phase curve are plotted in Fig. 5(c), the retrieved spectrum from the FROG signal exhibits multi-peak structures and agrees well with the output light spectrum directly measured using the spectrometer [see Fig. 5(e)]. The time-frequency hologram of the measured DW, retrieved from the FROG signal, is illustrated in Fig. 5(d). It was found that the multiple spectral components (peaks) correspond to, in the temporal domain, multiple DW packets emitted separately from the pump pulse. These measurements are in striking agreements with our simulations demonstrated above, perfectly verifying the mechanism of two-stage DW emission.

This cascaded DW emission process ensures a good coherence of the generated DW which has been verified by the smooth temporal phase curve of the measured DW, see Fig. 5(c). This coherent DW packet can be efficiently compressed using proper chirp compensation techniques. In the simulation, we found that the DW duration can be compressed to ~65 fs when only the second-order dispersion (-3852 fs$^2$) was compensated. Ultrashort DW pulse of ~28 fs duration (close to ~25 fs Fourier transform limit), could be obtained if the third-order (-99037 fs$^3$) and four-order (-1246460 fs$^4$) dispersion of the pulse were further compensated, see Fig. 5(f). Such high-order chirp compensation is practicable in the experiment through using some advanced chirp-compensation set-ups [31,32].

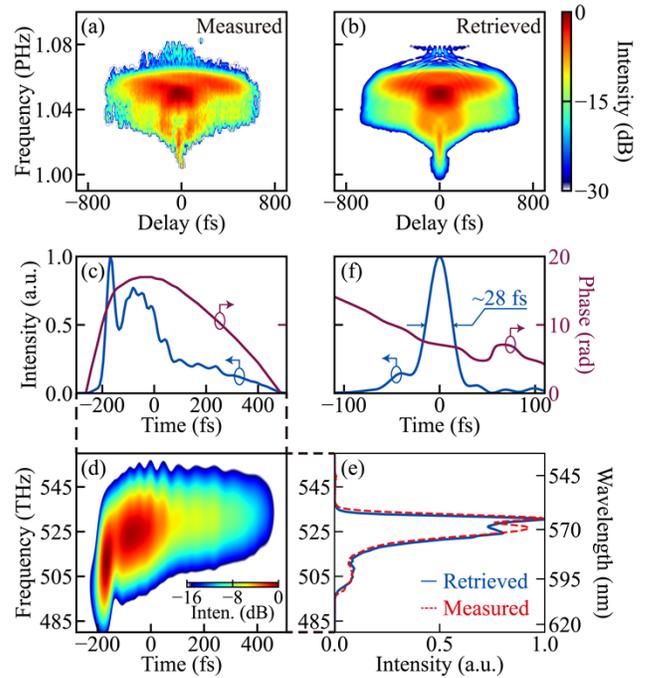

FIG. 5. Time-domain characterization of the multiple-peaks DWs, filtered out using the short-wavelength-pass filter with a cut-off wavelength of 600 nm, at input pulse energy of 3.9 μJ. (a) and (b) Measured and retrieved traces of SHG-FROG signal. (c) Retrieved temporal profile (solid blue line) and phase (solid purple line). (d) Time-frequency analysis of the measured DW using a 10-fs Gaussian gate pulse. (e) Retrieved spectral profile (solid blue line) and the reference spectrum (dashed red line)

directly measured using the spectrometer. (f) The temporal profile (solid blue line) and phase (solid purple line) of the compressed pulses after suitable dispersion compensation.

In conclusion, we studied both theoretically and experimentally the generation dynamics of broadband, multiple-peak DWs within a resonance band of Ar-filled AR-HCF. The delicate coupling mechanism between two-stage of soliton self-compression, soliton-plasma interactions, and phase-matched DW emission was unveiled, providing a few deep insights on the formation of structured DW spectra. The soliton blue-shifting effect due to soliton-plasma interactions, together with the DW phase-matching theory, can precisely predict the spectral positions of the multi-peak DW components. We found in the FROG measurements that the DW trace exhibited, in the temporal domain, two packets of quasi-linear waves with slightly-different central frequencies, successively shed from the pump pulse. The ultrashort DW light with specified central wavelength and high laser-conversion efficiency may have great application potentials in many experiments of ultrafast optics.

This work was supported by National Postdoctoral Program for Innovative Talents (Grant No. BX2021328), National Natural Science Foundation of China Youth Science Foundation Project (Grant No. 62205353), China Postdoctoral Science Foundation (Grant No. 2021M703325), Shanghai Science and Technology Innovation Action Plan (Grant No. 21ZR1482700), National High-level Talent Youth Project, Zhangjiang Laboratory Construction and Operation Project (Grant No. 20DZ2210300), National Natural Science Foundation of China (Grants No. 61925507 and No. 62275254).